\documentclass[modern]{aastex631}

\usepackage{upgreek}
\shorttitle{WR 125: the spectroscopic orbit and dust properties}
\shortauthors{Richardson et al.}
\graphicspath{{./}{figures/}}

\begin{document}

\title{The long-period spectroscopic orbit and dust creation in the Wolf-Rayet binary system WR 125}

\author[0000-0002-2806-9339]{Noel D. Richardson}
\affiliation{Department of Physics and Astronomy, Embry-Riddle Aeronautical University, 3700 Willow Creek Rd., Prescott, AZ 86301, USA}

\author[0000-0002-0786-7307]{Andrea R. Daly}
\affiliation{Department of Physics and Astronomy, Embry-Riddle Aeronautical University, 3700 Willow Creek Rd., Prescott, AZ 86301, USA}
\affiliation{Department of Physics \& Astronomy, University of Wyoming, 1000 E. University, Laramie, WY 82070, USA}

\author[0000-0002-8092-980X]{Peredur M. Williams}
\affiliation{Institute for Astronomy, University of Edinburgh, Royal Observatory, Edinburgh EH9 3HJ, United Kingdom}

\author[0000-0002-7648-9119]{Grant M. Hill}
\affiliation{W.M. Keck Observatory, 65-1120 Mamalahoa Highway, Kamuela, HI 96743, USA}

\author[0000-0002-2287-8151]{Victor I. Shenavrin}
\affiliation{Sternberg Astronomical Institute, Moscow State University, Universitetskij pr., 13, Moscow, 119991, Russia}

\author[0000-0002-9129-5988]{Izumi Endo}
\affiliation{Department of Astronomy, Graduate School of Science, The University of Tokyo, Bunkyo-ku, Tokyo 113-0033, Japan}

\author[0000-0002-1115-6559]{Andr\'e-Nicolas Chen\'e}
\affiliation{NSF's NOIRLab, 670 N. A'ohoku Place, Hilo, HI 96720, USA}

\author[0000-0003-3682-854X]{Nicole Karnath}
\affiliation{Space Science Institute, 4765 Walnut St, Suite B Boulder, CO 80301, USA}
\affiliation{Center for Astrophysics Harvard \& Smithsonian, Cambridge, MA 02138, USA}

\author[0000-0003-0778-0321]{Ryan M. Lau}
\affiliation{NSF's NOIRLab, 950 N. Cherry Avenue, Tucson, Arizona 85719, USA}

\author[0000-0002-4333-9755]{Anthony F. J. Moffat}
\affiliation{D\'epartement de physique, Universit\'e de Montr\'eal, Complexe des Sciences, 1375 Avenue Th\'er\`ese-Lavoie-Roux, Montr\'eal, Queb\'ec, H2V 0B3, Canada}

\author[0000-0001-9754-2233]{Gerd Weigelt}
\affiliation{Max Planck Institute for Radio Astronomy, Auf dem H\"ugel 69, 53121 Bonn, Germany}








\begin{abstract}

Several long-period binaries with a carbon-rich Wolf-Rayet star and an O star produce dust in their wind collisions. In eccentric binaries, this is seen most strongly near periastron passage. The exact conditions leading to dust creation require orbital properties to be determined, which is difficult owing to their long periods. Recently, the binary system WR\,125 (WC7+O9III) began a dust creation episode seen through an infrared outburst first detected by NEOWISE-R, which was the first outburst detected since 1991. We present new near- and mid-infrared photometry, which we use to show consistency between the two outbursts and derive an orbital period of 28.12$^{+0.10}_{-0.05}$ yr. We use a long time-series of optical spectra to place the first constraints on its orbital elements, on the assumption that this system will produce dust near periastron. The orbit has a mild eccentricity of 0.29$\pm$0.12 and is only derived for the Wolf-Rayet component, as the O star's radial velocities have noise that is likely larger than the expected semi-amplitude of the orbit. We also present SOFIA/FORCAST grism spectroscopy to examine the infrared {spectral energy distribution} (SED) of the dust during this outburst, comparing its properties to other WCd binaries, deriving a dust temperature of 580 K in 2021. This collection of observations will allow us to plan future observations of this system and place the system in the context of dust-creating Wolf-Rayet binaries. 

\end{abstract}



\section{Introduction} \label{sec:intro}


A classical Wolf-Rayet star is a massive, evolved star that has lost its outer envelope to reveal a compact, hydrogen-depleted, hot star with a strong stellar wind. These stars are usually either nitrogen-rich (spectral type WN) or carbon-rich (WC), and could form through either strong stellar winds \citep{1975MSRSL...9..193C} which would likely also include episodic mass loss \citep{2006ApJ...645L..45S}, or through binary interactions \citep[e.g.,][]{1998NewA....3..443V, 2009MNRAS.400L..20E}. The WC stars are sometimes observed to form dust, which is inferred through excess emission at infrared wavelengths first discovered by \citet{1972A&A....20..333A}. The dust production is often seen as persistent or periodic, but episodic bursts are also seen in some systems. Periodic and persistent dust makers are often seen in binary systems {as first discovered with} WR\,140 \citep[WC7+O5.5I;][]{1978MNRAS.185..467W} and WR\,48a \citep[WC9;][]{1983A&A...118..301D}. {Both of these systems} were observed to show rapid increases in the infrared attributed to dust formation. {WR 140 was subsequently shown to be a
high-eccentricity binary with dust formation near periastron
passage \citep{1990MNRAS.243..662W}. It is predicted that near the time of a periastron passage, the} density of wind collisions between the WC star and a companion OB star trigger dust formation. In contrast to the long-period episodically-producing dustar WR\,140, persistent dust makers tend to be in low eccentricity orbits. Curiously, some such systems like $\gamma^2$ Velorum have never been observed to form dust, although very similar to ones that do.

The dust production in WC binaries could be of cosmological importance. Both \citet{2007ASPC..367..213M} and \citet{2021ApJ...909..113L} suggest that the formation of WC binaries could be the first source of dust in a low metallicity environment. They can form dust earlier than supernovae because the higher mass and hence more rapidly evolving WC star can form through binary interactions before the supernova takes place, where the primary star loses its envelope through Roche lobe overflow onto its OB companion. Then, as the Roche lobe overflow ends, the strong stellar winds of the two stars will form a shock interface conducive to dust formation down stream from the heated apex.

Of the WC binary ``dustars", the prototype is often considered to be WR\,140. This system has a very well-established orbit, both spectroscopically \citep{2011MNRAS.418....2F,2021MNRAS.504.5221T} and with a visual orbit established through interferometry \citep{2011ApJ...742L...1M,2021MNRAS.504.5221T}. It has a long 7.93 yr period with high eccentricity ($e=0.8993\pm0.0013$) with the dust formation triggered by a changing gas density in the shock front near periastron. At this binary phase, the increase in the density of the shocked gas prompts increased X-ray production, but this seems to switch to cooling via optical emission lines at phases very close to periastron, perhaps allowing the conditions for dust production to occur \citep{2021ApJ...923..191P}.

\citet{2009MNRAS.395.1749W} found that the dust around this prototype ``dustar" (WR\,140) survived at least two cycles from imaging fossil dust around the system at mid-infrared wavelengths. The overarching question of dust survivability in these hostile environments was still open however in part due to limited spatial resolution and sensitivity with ground-based imaging, although \citet{2003ApJ...596.1295M} showed from ground-based mid-IR imagery that the dust in WR140 very likely reached the ISM. With a short exposure using \textit{JWST}+MIRI at mid-infrared wavelengths, \citet{2022NatAs...6.1308L} showed that the dust survives at distances out to 70,000 AU from WR\,140, likely implying that the dust should survive and be included in the dust budget of galaxies.

WR\,125 is a near spectroscopic twin to WR\,140 \citep{1986ApJ...303..239A}, and is the topic of this paper, first noted as IC 14-36 by \citet{1956BOTT....2n..31I}. \citet{1987A&A...182...91W} observed this star with infrared photometry but found it not to show variation in its infrared flux. However, its near twin status prompted continued infrared monitoring until \citet{1992MNRAS.258..461W} discovered an infrared outburst in the years 1990--1991. \citet{1994MNRAS.266..247W} reported on additional infrared photometry of WR\,125 during this outburst and presented some of the first optical spectroscopy of the system. They confirmed that the emission line dilution is caused by an O9III companion star based on the optical absorption lines present. Furthermore, \citet{1994MNRAS.266..247W} presented infrared spectroscopy of the system and the dust cloud, finding no signatures of the WR wind lines in the 10 $\upmu$m region, nor the 11.52 $\upmu$m graphite feature, suggesting that the dust had to be amorphous carbon.

Very little progress was made on WR\,125 for many years following this initial discovery of dust creation and that the star had remarkable similarities to WR\,140. While the star had been occasionally included in infrared sky surveys, infrared monitoring was largely absent. The launch of the Wide-field Infrared Survey Explorer (WISE) in 2009, which was reactivated \citep{2014ApJ...792...30M} as the Near-Earth Object Wide-field Infrared Survey Explorer (NEOWISE-R) in 2013 December following a hibernation of the satellite that began in 2011 February. While NEOWISE-R was unable to observe at the longer wavelengths from the WISE mission, its continued sky monitoring at 3.4 and 4.6 $\upmu$m showed that the WR\,125 system began a second dust creation outburst in 2018 as reported by \citet{2019MNRAS.488.1282W}, indicating that the binary was once again approaching the phases amenable to dust production.

The increase in infrared flux observed with the NEOWISE-R mission \citep{2019MNRAS.488.1282W} has prompted a few new studies into WR\,125. 
\citet{2019MNRAS.484.2229M} reported on a few public X-ray observations of WR\,125 taken with the \textit{Swift} and \textit{XMM-Newton} X-ray observatories. The observations, largely taken prior to the start of the current dusty outburst, were fairly constant in their flux and shape. The flux was the same as observed in 1981 with the \textit{Einstein} satellite, but the flux was lower during the previous dust outburst in 1991 as observed with \textit{ROSAT}. This is likely similar to that of the prototype of the dust-making WC binaries, WR\,140, which also shows an X-ray dip at some phases close to periastron as described by \citet{2021ApJ...923..191P}.

In addition to the X-ray observations, \citet{2022ApJ...930..116E} presented a mid-infrared spectrum of the system taken with Subaru and the COMICS instrument. They found that the system shows a broad 8 $\upmu$m feature that was also seen in several WCd stars observed with \textit{ISO} and the SWS instrument. This feature is a typical ``unidentified infrared" (UIR) band, which often correlates to other infrared features. These UIR features are also seen in spatially-resolved spectroscopy of the dust surrounding WR\,140 and imaged with \textit{JWST} \citep{2022NatAs...6.1308L}. \citet{2022ApJ...930..116E} presented MIR spectroscopic observations {of WR\,125} obtained in 2019 that were characterized with a blackbody temperature of nearly 800 K after removing the underlying stellar flux. This was hotter dust than seen in the other WR binaries they used as a comparison from \textit{ISO} observations, likely as it was more recently formed. They also derived a period of $\sim$28.1 yr by a comparison of a single infrared flux point with the light curve presented by \citet{1992MNRAS.258..461W,1994MNRAS.266..247W}.

Many of these results show that WR\,125 is a prime target for additional observations and understanding the dust production around these systems. To date, no spectroscopic orbit has been measured, and a time-series analysis of the infrared photometry is strongly needed. To that end, we present new observations in Section 2 and analyze the time-series photometry in Section 3 to obtain a period with higher confidence than elsewhere. In Section 4, we present the first spectroscopic orbit of the system. We present an infrared spectrum taken with SOFIA and FORCAST in Section 5. We discuss these findings in Section 6, and then conclude our study. 


\section{Observations} \label{sec:obs}

\subsection{Infrared Photometry}
Near- and mid-infrared photometry of WR\,125 has been collected with the Sternberg Astronomical Institute’s (SAI) Crimean Laboratory of Moscow State University using a $JHKLM$ photometer with an InSb photovoltaic detector cooled with liquid nitrogen \citep{2011ARep...55...31S}. Observations have been taken regularly since the beginning of the current dust creation episode that was observed with NEOWISE-R \citep{2019MNRAS.488.1282W} and are shown in Fig.~\ref{fig:all phot}; subsequent observations of WR\,125 with NEOWISE-R are severely saturated. The new data allow comparison with the previously observed outburst. One of these observations was also used by \citet{2022ApJ...930..116E}. Each data point consists of multiple sub-exposures of 30--60 s with a total integration time of 5–10 minutes in each of the $JHKL$ filters and 20--25 minutes in the $M$ filter. The observations are compared to the standard star BS 7488 \citep{1966CoLPL...4...99J} observations taken before or after each observation of the target. The light curve is tabulated as an online data file. In addition to these new data, along with the archival measurements from \citet{1992MNRAS.258..461W,1994MNRAS.266..247W}, we include the two measurements from 2007 and 2008 taken at UKIRT and reported by \citet{2022ApJ...930..116E} in this analysis. {For our analysis, we have combined $L$ and $L'$ data into one light curve, as this is also how archival data \citep[e.g.,][]{1994MNRAS.266..247W} have been presented.}

\subsection{Optical Spectroscopy} 
We have collected spectroscopy of WR\,125 with three instruments. These spectra typically have a signal-to-noise of at least 100 at wavelengths longer than $\sim 5000$\AA\ and were all reduced using standard pipelines and techniques for the instruments used. The first set of spectra were obtained with the Keck I telescope and the Low Resolution Imaging Spectrometer \citep[LRIS;][]{1995PASP..107..375O, 2010SPIE.7735E..0RR} on seven independent nights, which provides spectra with wavelength coverage from $\sim4900$\AA\ to $\sim6200$\AA. The spectrograph had a dispersion of 0.64 \AA\ pixel$^{-1}$ and we typically obtained a signal-to-noise of 100 per sub-exposure with a 5-minute exposure. 

We also collected spectra with the Keck II telescope and the Echellette Spectrograph and Imager \citep[ESI; ][]{2002PASP..114..851S}. These data have a resolving power of 13,000 and cover several orders of the spectrum between 3900 \AA\ and 1.1 $\upmu$m. Near the \ion{C}{3} $\lambda$5696 line, our data had a typical signal-to-noise of 100 in an order of the spectrograph covering 5100 to 5970 \AA\ with a dispersion of 0.21 \AA\ pixel$^{-1}$. A typical spectrum was obtained {with a 5-minute exposure}.  

Lastly, we obtained a few spectra with Gemini-North and the Gemini Multi-Object Spectrograph \citep[GMOS; ][]{2004PASP..116..425H}. We used the B600 grating with the G5307 blocking filter to obtain spectra between 4115 \AA\ and 7410 \AA, with a dispersion of 1.0 \AA\ pixel$^{-1}$. In the vicinity of the \ion{C}{3} $\lambda$5696 line, our data had a typical signal-to-noise of 100 for an exposure of 100 s. 

\subsection{SOFIA Infrared Spectroscopy}

We obtained mid-IR grism spectroscopy using the FORCAST instrument \citep{2013PASP..125.1393H, 2018JAI.....740005H} onboard SOFIA \citep{2018JAI.....740011T} during the peak of the recent dust creation outburst. 
These observations, taken on 2021 April 8, used both the G063 and G111 grating setups. The G063 setup covers the wavelength range of 4.9--8.0 $\upmu$m, but we found the spectra near the edge of the chip to be too noisy for analysis, effectively reducing the useful range to 5.1--7.9 $\upmu$m. The G111 setup was used for a spectrum in the range of 4.9--8.0 $\upmu$m. Both of these spectra were taken with the 2.4$\arcsec$ slit, yielding a spectral resolving power of $\sim 180$ and $\sim 260$ respectively. There was an additional spectrum taken with the G227 setup that covers the 17.6-–27.7 $\upmu$m range. This setup provided very low signal-to-noise, so we use this primarily {to provide} a flux point at 23 $\upmu$m. All data were processed through the typical SOFIA pipelines.
The FORCAST grism data are known to have variable slit-losses, adding uncertainty to the flux calibration, but these data should be accurate to a few percent \citep[e.g.,][]{2021ATel14794....1G}.

\section{The Orbital Period from the Infrared Photometry} \label{sec:period}

\begin{figure}
    \centering
    \includegraphics[width=0.7\columnwidth]{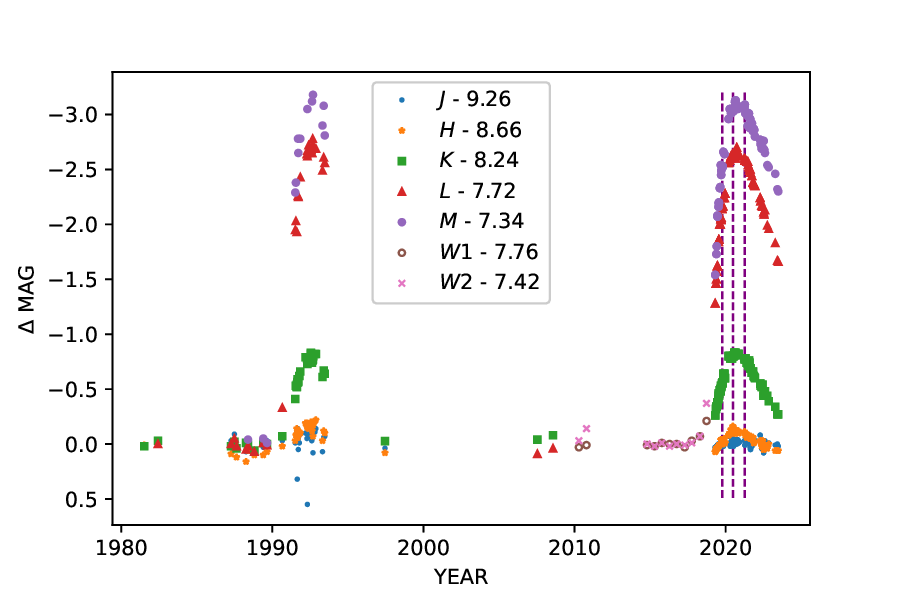}
    \caption{Infrared photometry from \citet{1992MNRAS.258..461W,1994MNRAS.266..247W} as well as new measurements of WR\,125. NEOWISE-R $W1$ and $W2$ after 2018 are too heavily saturated to use. We have also incorporated measurements from 2MASS. For each bandpass, we have shifted the points to the same quiescent level to highlight the outbursts and different amplitudes as a function of wavelength. The new observations are tabulated in a machine-readable format as data behind the figure. The vertical lines indicate the time of the infrared spectrum reported by \citet{2022ApJ...930..116E} in 2019, the 2020 spectrum reported by Endo (2022, PhD diss) and our SOFIA spectrum obtained in 2021. }
    \label{fig:all phot}
\end{figure}

Fig.~\ref{fig:all phot} shows the infrared light curve of WR\,125. In this figure, we see that the near-infrared $J$-band flux does not show much excess during the two outbursts, while $H$-band shows a slight excess. The $K$-band flux reaches nearly a magnitude of excess during an outburst, while $L$/$L'$-band and $M$-band reach about 2.5 and 3 magnitudes of excess compared to the quiescent flux. We note that the amplitude of the $L$/$L'$ and $W1$ outbursts are heterogeneous, but the amplitude of variation at these wavelengths is so {large}
that we have not attempted to put the data on a common scale. The addition of the two points in 2007-2008 from UKIRT and knowing the $K$-band magnitude reported in 2MASS \citep[taken between 1997.5 and 2001.1][]{2006AJ....131.1163S} was 8.214$\pm$0.017, provides enough information to know that these two recorded outbursts were consecutive outbursts for the system despite poor coverage in the time-series as they rule out the presence of another outbust
mid-way between the two observed outbursts.

\begin{figure}
    \centering
    \includegraphics[angle=90,width=0.7\columnwidth]{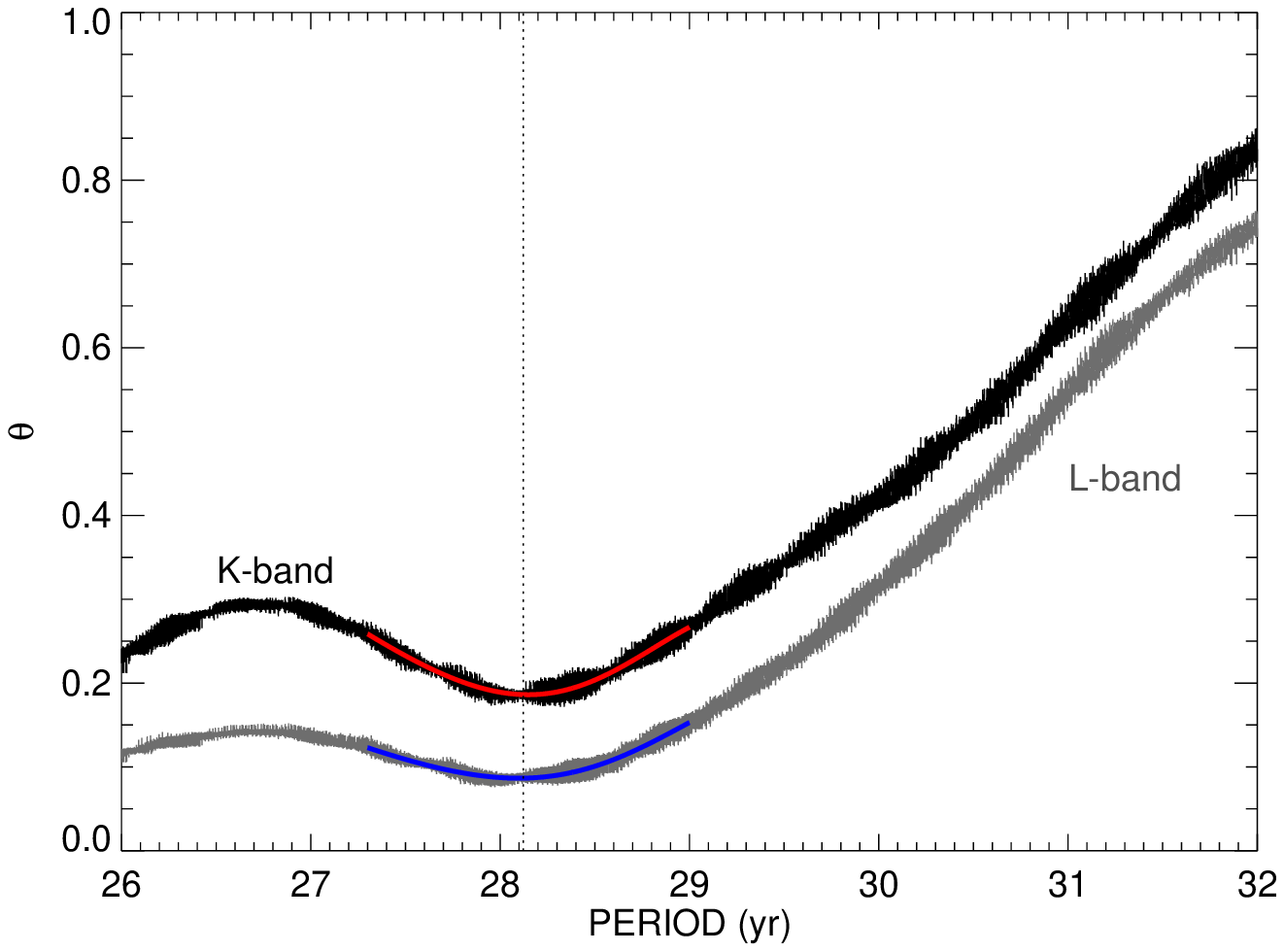}
    \caption{Results of the phase dispersion minimization for determination of the period based on the $K$- and $L$/$L'$-band photometry. The resulting period is 28.12 years.}
    \label{fig:pdm}
\end{figure}

Given the nature of WR\,125, we can use the infrared light curve to derive a binary period. We began our time-series analysis with Fourier techniques using {\tt Period04} \citep{2005CoAst.146...53L}. Fourier techniques assume a sinusoidal function, which is not the observed shape of the light curve we see in Fig.~\ref{fig:all phot}, so the results tended to lie on harmonics of the fundamental (orbital) period that is seen to be $\sim28.1$ yr \citep{2022ApJ...930..116E}. Determining the true period based on harmonics introduces extra errors because of the long time-scales involved. As we wish to use this period for a spectroscopic orbit, we needed a different approach. 

We then used the phase dispersion minimization routines described by \citet{1978ApJ...224..953S} to determine the period independent of the light curve shape. This method calculates a dispersion statistic at each potential period, and a minimum in this quantity represents a candidate period. We show this in Fig.~\ref{fig:pdm}, where we see a minimum at presumed period at 28.12$^{+0.10}_{-0.05}$ years, as well as the harmonic at 2/3 the period (3/2 the frequency) of $\sim18$ years. {The error on this period comes from the statistical uncertainty in the pdm statistic. }We calculated this statistic for the $K$- and $L$-band light curves. The shorter-wavelength bandpasses had a smaller variability amplitude, casting doubt on the minimum found, while the longer-wavelength $M$-band has a sparser light curve for the calculation. 

\begin{figure}
    \centering
    \includegraphics[width=0.7\columnwidth]{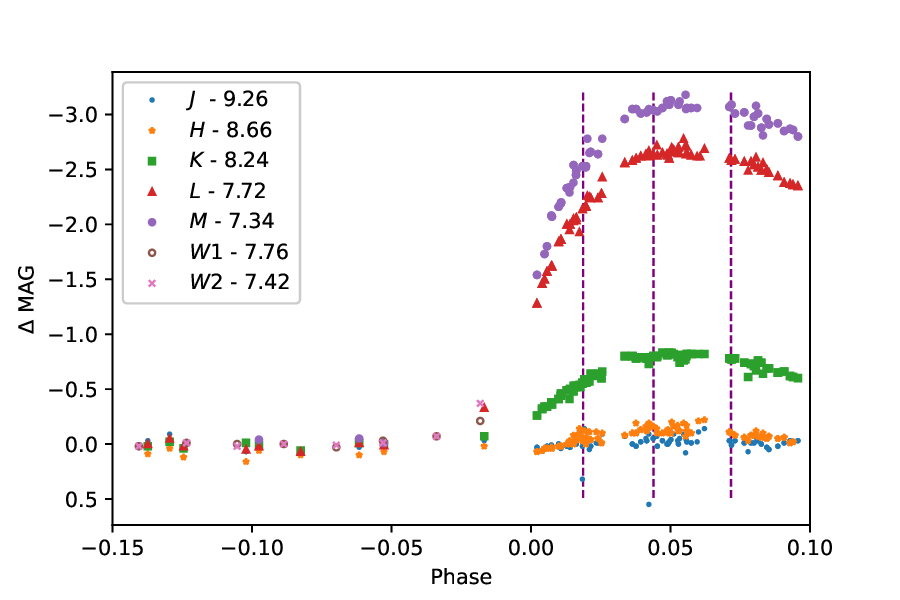}
    \caption{Phased infrared light curves based on the phase-dispersion minimization routines used in Section \ref{sec:period}. We show photometry from both outbursts here, illustrating the repeatability of the photometry from one cycle to the next. As in Fig.~\ref{fig:all phot}, the vertical lines represent the timing of the three mid-infrared spectra: from \citet{2022ApJ...930..116E} in 2019, Endo (PhD dissertaion, 2022) in 2020, and then our SOFIA spectrum obtained in 2021. }
    \label{fig:Phase}
\end{figure}

From the $K$- and $L$-band light curve analyses, we derive a period of 28.12$^{+0.10}_{-0.05}$ years by fitting the minima with Gaussian curves, overlaid as red and blue curves in Fig.~\ref{fig:pdm}. This period is consistent with other recent papers \citep{2021AJ....162..257A, 2022ApJ...930..116E}, and we show the phased light curves in Fig.~\ref{fig:Phase}. In the phased light curves, we assume that the outburst begins as the {\it WISE} flux began going into an excess, and that the plateau in the $K$-band peak represents the end of new dust production. We then take the mid-point time between these two epochs to be the time of a periastron passage, assuming that the dust-creation event should be centered on the periastron passage. We checked this period with the Lafler-Kinman approach for sparse time-series \citep[e.g.,][]{2017AJ....154..231S} and found the same result.

\section{Orbital Elements} \label{sec:orbit}

\begin{figure}
    \centering
    \includegraphics[angle=90,width=6in]{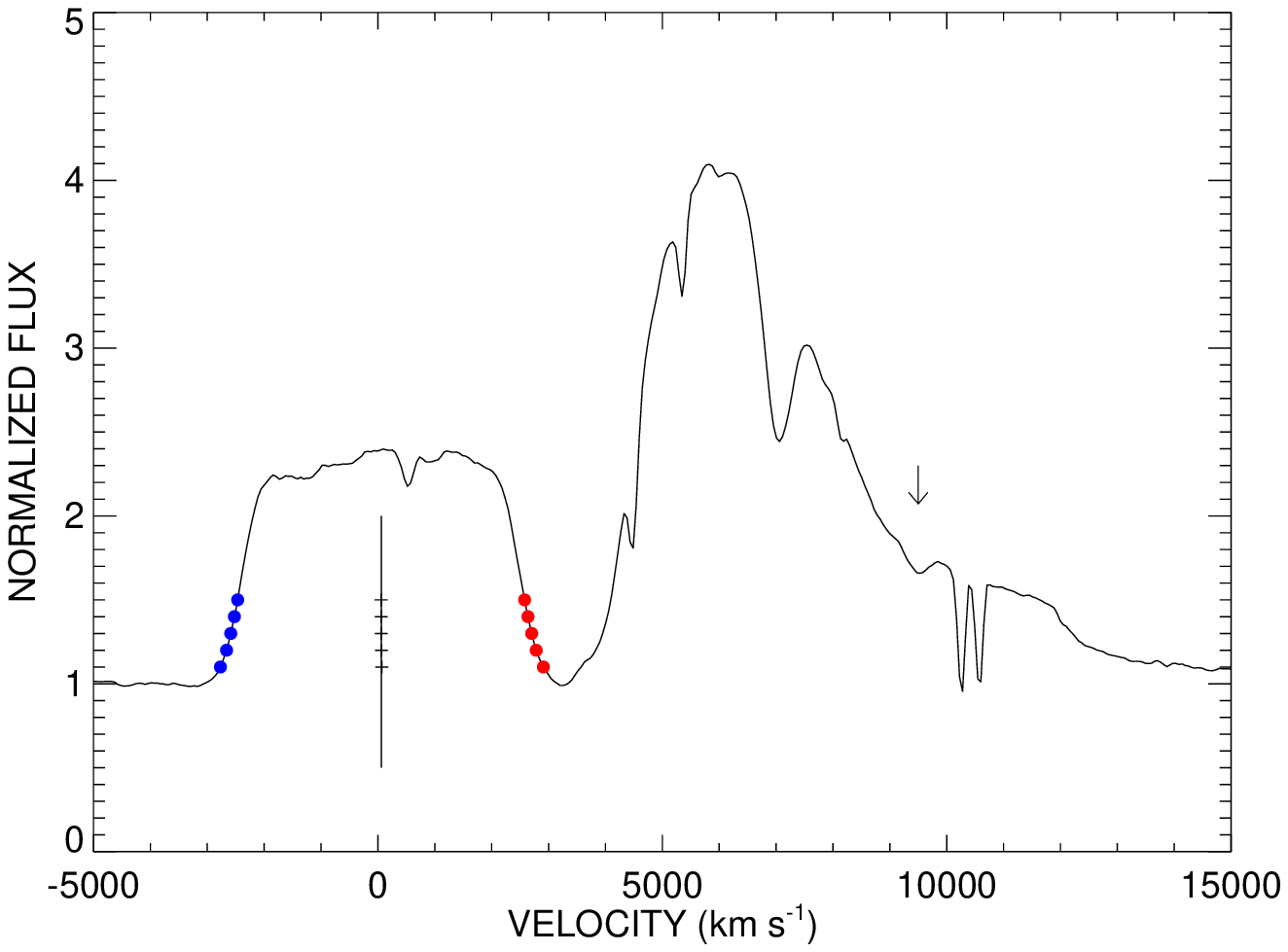}
    \caption{An example spectrum of WR\,125 in the region around \ion{C}{3} $\lambda$5696 (reference for the velocity scale) from Gemini/GMOS on UT 2020 July 7. We show the measured blue and red measured velocities on the line wings with cross hairs on the central line indicating the measured velocity. The small absorption at velocity $+9500$ km s$^{-1}$ is the \ion{He}{1} $\lambda$5876 from the O star's photosphere, which we point out with a vertical arrow.}
    \label{fig:spec}
\end{figure}

In order to measure the radial velocities of WR\,125, we employed similar techniques as that done in the WR\,140 system by \citet{2021MNRAS.504.5221T}. We examined the fairly isolated \ion{C}{3} $\lambda$5696 emission line to measure radial velocities of the WR star. We examined the normalization of each spectrum in the region surrounding the emission line and applied local corrections as necessary. Unlike WR\,140, we did not see any strong evidence of excess emission in the \ion{C}{3} line although \citet{1992MNRAS.258..461W} observed some excess red emission in the \ion{He}{1} $\lambda$10830 line. We measured the line using bisectors at five points above the continuum level. A radial velocity measurement was taken to be the average of these five points, with a standard deviation of the individual points used as an error measurement, and are shown in Table \ref{rvs} and illustrated in Fig.~\ref{fig:spec}. For our data taken with LRIS, we often had several sub-exposures which we measured independently to verify consistency in our measurements.

\begin{deluxetable}{lccc}
\tablewidth{0pc} 
\tablecaption{Wolf-Rayet velocities\label{rvs}} 
\tablehead{ 
\colhead{HJD}                      & \colhead{Radial Velocity}  & \colhead{(O -- C)} & \colhead{Instrument} \\
\colhead{-2,450,000}    & \colhead{(km s$^{-1}$)}  & \colhead{(km s$^{-1}$)} & \colhead{}} 
\startdata 
       2852.9574  &  70.0$\pm$9.1 & -3.5 &  LRIS \\
       2852.9601  &  70.7$\pm$8.3 & -2.8 &  LRIS \\
       2852.9628  &  69.7$\pm$6.8 & -3.7 &  LRIS \\
       2875.9852  &  64.7$\pm$9.1 & -8.6 &  LRIS \\
       2876.0080  &  85.4$\pm$10.2 & 12.1 & LRIS \\
       2876.0128  &  83.1$\pm$7.8 &  9.8 & LRIS \\
       2876.0208  &  78.8$\pm$11.4 & 5.5 &  LRIS \\
       3179.0848  &  72.3$\pm$12.8 & 1.9 &  LRIS \\
       3179.1020  &  75.1$\pm$8.7 &  4.7 & LRIS \\
       3202.0537  &  65.5$\pm$10.4 & -4.7 &  LRIS \\
       3202.0582  &  65.5$\pm$4.2 & -4.7 &  LRIS \\
       3202.0627  &  70.7$\pm$1.9 &  0.5 & LRIS \\
       4273.1227  &  62.9$\pm$10.8 &  4.7 & LRIS \\
       4273.1251  &  63.4$\pm$4.9 &  5.1 & LRIS \\
       4381.8594  &  36.7$\pm$6.3 & -20.2 &  LRIS \\ 
       4381.8639  &  41.2$\pm$8.8 &  -15.7 & LRIS \\
       4381.8683  &  40.0$\pm$9.8 &  -17.0 & LRIS \\
       4646.1210  &  68.1$\pm$4.7 &  14.6 & LRIS \\
       4646.1247  &  71.0$\pm$8.1 &  17.4 &  LRIS \\
       6516.9958  &  30.3$\pm$6.7 &  3.2 & ESI \\
       6517.0040  &  30.3$\pm$7.9 & 3.1 &  ESI \\
       7219.0433  &  24.9$\pm$3.9 & 5.8 &  ESI \\
       7626.8504  &   6.4$\pm$2.6 & -11.4 &  ESI \\
       8777.7180  &  59.3$\pm$1.6 & 16.6 &  GMOS \\
       8809.6989  &  45.7$\pm$1.9 &  1.9 & GMOS \\
       9010.1081  &  31.1$\pm$10.1 & -20.2 &  GMOS \\
       9019.0753  &  31.5$\pm$2.6 & -20.2 &  GMOS \\
       9023.0231  &  37.0$\pm$3.3 &  -14.8 & GMOS \\
       9023.9995  &  45.9$\pm$6.5 &  -5.9 & ESI \\
       9036.0122  &  52.8$\pm$1.1 &  0.6 & GMOS \\
       9038.0196  &  60.6$\pm$2.8 &  8.2 & GMOS \\
       9371.0764  &  69.4$\pm$2.4 & 5.7 &  ESI \\
       9814.8366  &  45.0$\pm$5.7 & -30.0 & ESI \\
       10150.0050 &  56.5$\pm$10.2 & -24.0 & ESI \\ 
\enddata 
\end{deluxetable}

We attempted to measure the O star's absorption line velocities using the \ion{He}{1} $\lambda$5876 line, but found that this did not produce realistic measurements that would show an orbit with time. This line is seen in emission in late sub-type WC stars and it is possible that this occurs in the spectrum of WR\,125, making interpretation difficult. Our spectra did not extend to the higher Balmer absorption lines attributed to the companion \citep{1994MNRAS.266..247W}.

Given the sparse nature of the spectroscopic measurements, we found that fitting an orbit was not a straight-forward task. We utilized the orbit-fitting procedures used for the $\gamma^2$ Velorum system \citep{2017MNRAS.471.2715R}. The RV data span about 20 years so we did not search for a period in the RV data. Spectroscopy over at least another decade will be required to cover the dust-formation cycle and longer to establish a more definitive period. Meanwhile, therefore, by analogy with WR\,140 and WR\,137 whose dust-formation and spectroscopic periods are the same,  we fixed the period of WR\,125 to 28.12 years based on the infrared light curves (see previous section). 

\begin{figure}
    \centering
    \includegraphics[angle=90,width=0.7\columnwidth]{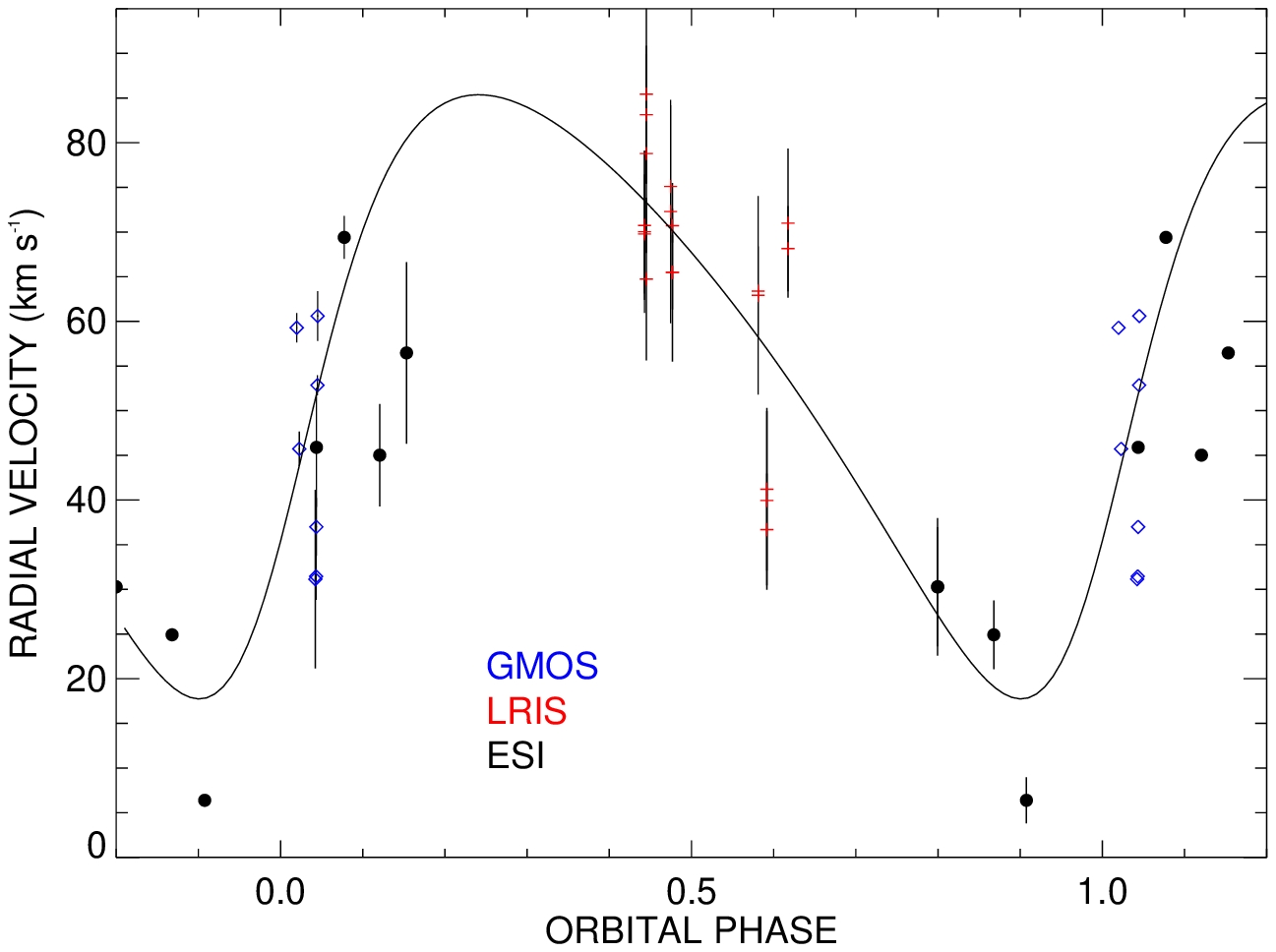}
    \caption{The single-lined spectroscopic orbit representing the motion of the WR star in the WR\,125 binary. The data are represented with black circles for Keck/ESI velocities, blue diamonds for Gemini-N/GMOS, and red $+$ signs for Keck/LRIS.}
    \label{fig:orbit}
\end{figure}

Similarly, to estimate the date of periastron passage, we look to the infrared light curve. It is evident that the dust condensation by WR\,125 occurred over a relatively longer interval than that of WR\,140 but in both cases the only way the wind-collision region could ``know'' the orbital phase is through the variation of the ambient pre-shock wind density and stellar radiation at the collision, which vary as the inverse square of the stellar separation. The conditions necessary for dust condensation must occur at critical stellar separations, and likely symmetrical in phase about periastron passage, as demonstrated in the case of WR\,140 by \cite{2022Natur.610..269H}. For the present we therefore assume that periastron passage occurs between the beginning and end of dust condensation. To that end, we used the first observation with NEOWISE-R as the system began to brighten to estimate the onset of dust production \citep{2019MNRAS.488.1282W}. We estimate that the dust production ends when the infrared light curve reaches a maximum in the near-infrared $K$-band, after which dust being carried away by the stellar winds is no longer being replenished by the condensation of new dust. We then estimate that the midpoint of these two times was the periastron passage to fit an orbit. The result of this fit is shown in Fig.~\ref{fig:orbit}, with the orbital elements given in Table \ref{orbital elements}. The resulting orbit has a moderate eccentricity, but will need better spectroscopic coverage in the coming decades to be refined.

\begin{deluxetable}{lc} 
\tablewidth{0pc} 
\tablecaption{Orbital Elements of the WC component star\label{orbital elements}} 
\tablehead{ 
\colhead{Element}                      & \colhead{Value}       } 
\startdata 
$P$ (d)                       \dotfill & 10,272\tablenotemark{a} \\ 
$T$ (HJD - 2,400,000)        \dotfill & $58575.5$\tablenotemark{b}     \\ 
$e$                           \dotfill & $0.29 \pm 0.12$          \\
$\omega$ (deg)                \dotfill & $241.5 \pm 5.1$             \\
$K_1$ (km s$^{-1}$)           \dotfill & $33.8 \pm 4.6$         \\ 
$V_0$ (km s$^{-1}$)           \dotfill & $56.2 \pm 2.1$          \\ 
$a_1\sin i$ ($R_\odot$)       \dotfill & $6600 \pm 900$           \\ 
$f(M)$ ($M_\odot$)            \dotfill & $36 \pm 15$            \\ 
r.m.s. (km s$^{-1}$)          \dotfill & 11.1                    \\ 
\enddata 
\tablenotetext{a}{Fixed to value obtained from analysis of infrared photometry (\S 3).} 
\tablenotetext{b}{Assumed to be the midpoint of the infrared rise.} 

\end{deluxetable} 


\section{The mid-infrared spectrum as observed by SOFIA} \label{sec:sofia}

\begin{figure}
    \centering
    \includegraphics[width=6in]{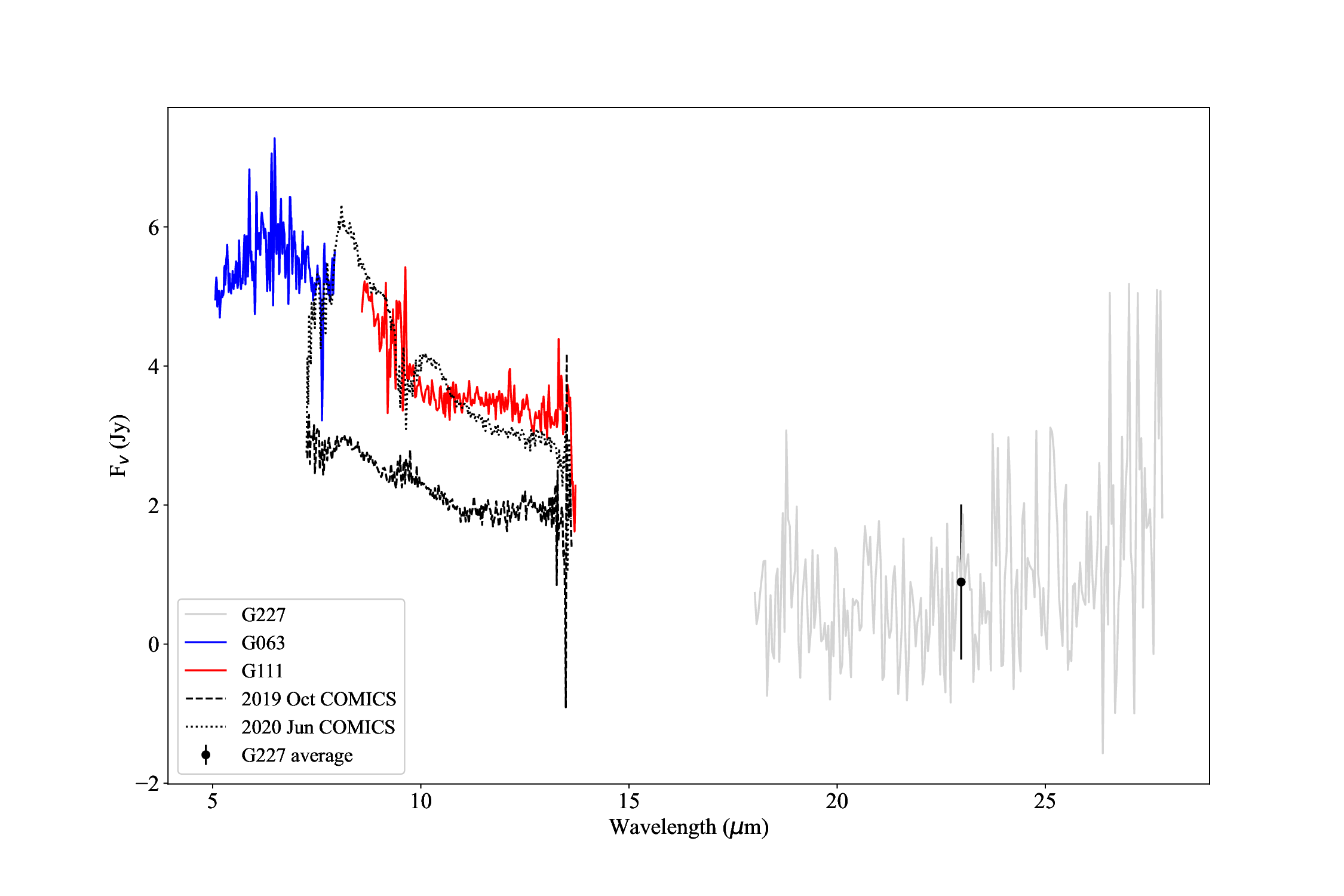}
    \includegraphics[width=6in]{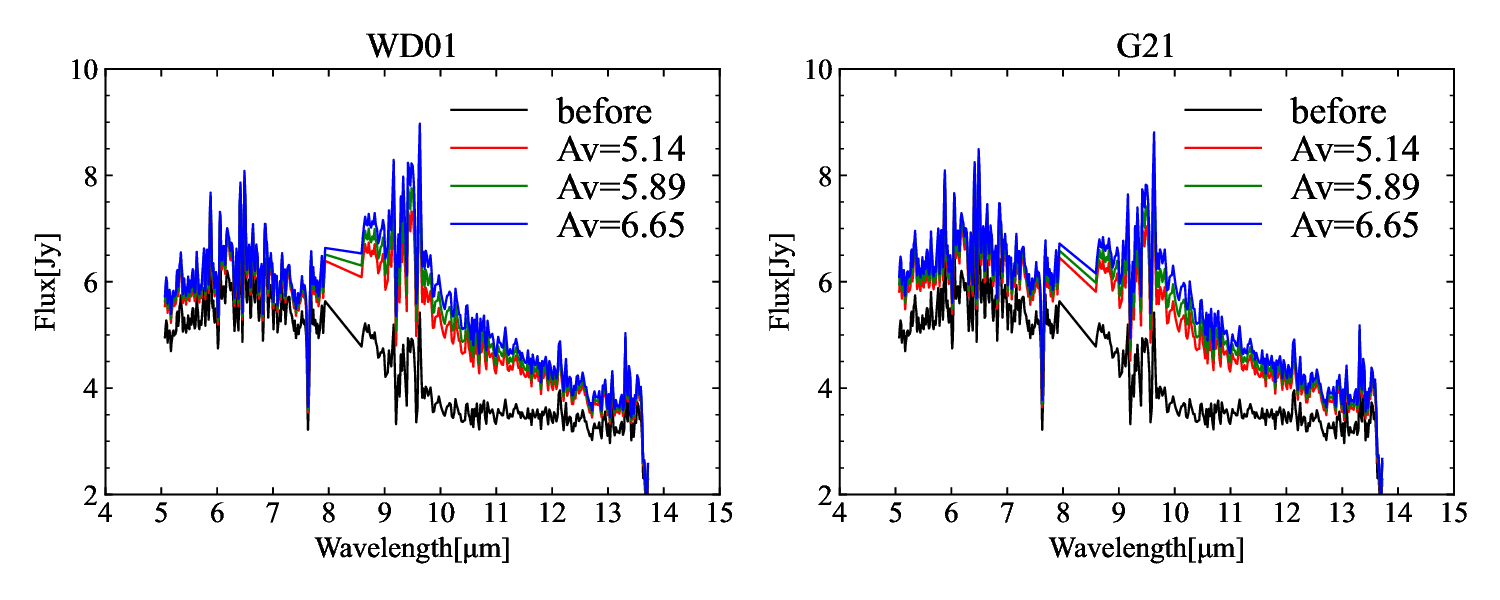}
    \caption{The flux-calibrated SOFIA data taken in 2021 April, compared to the spectrum presented by \citet{2022ApJ...930..116E} and a follow-up Subaru/COMIC spectrum presented by Endo (2022, PhD dissertation). Each grism setting from FORCAST is shown in a different color, and we also show the averaged flux of the G227 grating as a dark point with the standard deviation of all flux-calibrated points for this observation overplotted. In the second set of plots, we show the extinction-corrected flux of the SOFIA spectra for the values of $A_V$ discussed by \citet{2022ApJ...930..116E}, with the range corresponding to the error in $A_V$.}
    \label{fig:sofia}
\end{figure}

The spectrum of WR\,125 observed by SOFIA is shown in Fig.~\ref{fig:sofia}. The three grism spectra are of moderate signal-to-noise. The raw flux shows a peak near 6--7 $\upmu$m, and the data taken from 17.6–27.7 $\upmu$m with the G227 grating are too noisy for a thorough analysis given the lower flux and sensitivity of the instrument at these wavelengths. Following the procedures used by \citet{2022ApJ...930..116E}, we aimed to correct the fluxes for interstellar extinction using the extinction laws from \citet{2001ApJ...548..296W} (WD01) and \citet{2021ApJ...916...33G} (G21). \citet{2022ApJ...930..116E} noted that the value of $A_V$ was calculated to be $5.89\pm0.75$ from $A_v = 6.48\pm0.83$ by \citet{2020MNRAS.493.1512R} from the relation $A_v = 1.1 A_V$ \citep{1982IAUS...99...57T}. We show the extinction correction for $A_V = 5.89$ as well as the values at the extrema of the value reported by \citet{2020MNRAS.493.1512R} in Fig.~\ref{fig:sofia}.

\begin{figure}
    \centering
    \includegraphics[width=5in]{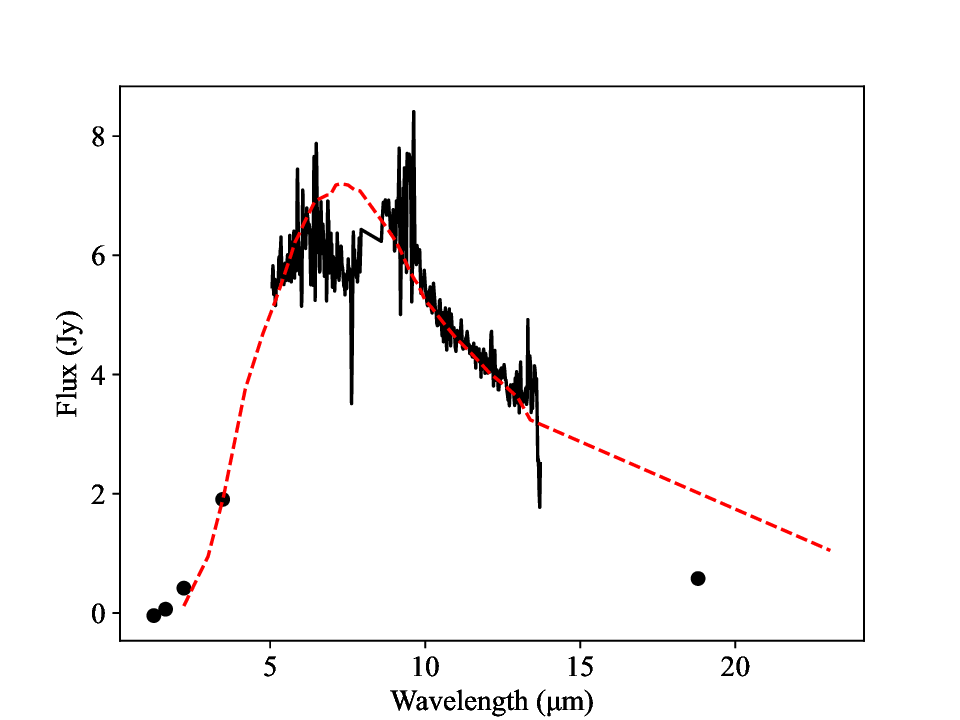}
    \caption{{\color{blue}After correcting the infrared SOFIA data for interstellar extinction (WD01 correction, $A_V = 5.89$) and subtracting the stellar contribution as approximated by a power-law, we fit a blackbody distribution to the SOFIA data and SAI photometry including the emissivity of the dust and found the temperature of the dust to be $\sim500$ K as shown in the red dashed line. A warmer or cooler temperature can change the fit to better compare to the short- or long-wavelength ends of the distribution, but this is similar to the other fits we examined.}}
    \label{fig:blackbody}
\end{figure}

We also used the photometry obtained at SAI converted to flux\footnote{We used the tools at https://irsa.ipac.caltech.edu/data/SPITZER/docs/dataanalysistools/tools/pet/magtojy/.} in our analysis of the SOFIA data. We began our analysis by subtracting the stellar contribution that was calculated by \citet{2022ApJ...930..116E}. The stellar contribution was calculated as a power-law with two forms based on the extinction law adopted. The power-law fit held the form of $$a \times \lambda^{-b},$$
where the coefficients were derived from dust-free epochs of infrared photometry and correspond to $(a,b) = (2.08,1.58)$ for the WD01 extinction law and $(a,b) = (1.79, 1.40)$ for the G21 extinction law. In both cases, the units of this relationship provide the flux in Jy and use a wavelength given in $\upmu$m and was fit with the infrared flux observed in the 1980s when no dust was seen in the SED. The $J$-band light curve shows little or no variability, so this relationship will provide the normal $J$-band flux. In both extinction situations, a blackbody fit provides a temperature for the dust of 400 K (Fig.~\ref{fig:blackbody}) {if we neglect the emissivity of the dust}. This is cooler than the temperature derived by \citet{2022ApJ...930..116E}, but also was from data taken two years later when the dust cloud should have expanded and cooled and is past the time of active dust formation as the $KLM$ light curves were dimming at the time of the SOFIA observations (see Fig.~\ref{fig:Phase}).

The SOFIA spectra were relatively short exposures and thus have limited signal-to-noise. However, there is a small rise at the long-wavelength end of the G063 spectrum and a general downward slope at the short-wavelength end of the G111 spectrum. We think that these spectral features are part of the 8 $\upmu$m feature reported by \citet{2022ApJ...930..116E}. This feature is often associated with a feature near 6-6.5 $\upmu$m. We see a noisy feature that is reminiscent of the feature present in the dust of WR\,137 that is discussed by \citet{2023arXiv230811798P}. Unlike the SOFIA data for WR\,137, we see a shorter wavelength of the feature indicating a higher hydrogen-content than that of WR\,137. The WR wind lines observed in WR\,137 early in its dust creation episode \citep{2023arXiv230811798P} are not easily seen in the spectrum of WR\,125 likely due to the larger dust contribution for WR\,125.

\section{Discussion} \label{sec:discuss}

WR\,125 is a member of the unusual class of dust-forming carbon-rich binaries. It has spectral properties similar to the nearby and well-studied prototype of these systems, WR\,140. The O star in the WR\,125 system is a later type and less luminous than in WR\,140, but the spectrum does not show a disk-like geometry around the O star like in the WR\,137 system \citep{2020MNRAS.497.4448S}. Given that the dense disk around the O star in WR\,137 may help promote the dust production, WR\,125 represents a longer period system with stellar parameters closer to that of WR\,140. 

Our derived orbit of WR\,125 shows a moderate eccentricity compared to the extreme value for WR\,140. WR\,140 has an eccentricity of 0.8993$\pm$0.0013 \citep{2021MNRAS.504.5221T}. In contrast, the unusual WR\,137 system, while also possessing a WC7 primary star has a very low eccentricity of 0.178 according to the spectroscopic orbit of \citet{2005MNRAS.360..141L}, although the decretion disk around the Oe star likely adds to the dust production \citep{2023arXiv230811798P} and the narrow geometry of the dust plume \citep{2023arXiv231115948L}. We also note that a new visual orbit of WR\,137 (Richardson et al.~in prep) reports a larger eccentricity of 0.314$\pm$0.001 and casts additional doubts on the O star radial velocities of \citet{2005MNRAS.360..141L}.

Recently, WCd binary systems have received much attention due to their photogenic appearance in imaging from the recent JWST observations first presented by \citet{2022NatAs...6.1308L}. The imaging-based geometries can be reconciled with models for dust production based on orbital elements and the balance of momenta from the stellar winds. With the orbital kinematics for the WR star derived for WR\,125, future observations of the dust from JWST will be able to create models for the dust production with fewer free parameters. 

With the long period for WR\,125, we can also consider the types of observations that could better pin down the orbit and evolution of the system. Despite the $\sim 8$ yr period of WR\,140, \citet{2021MNRAS.504.5221T} presented a binary evolution model for the system that showed that the stars likely interacted to produce the modern-day WR star and stellar masses we observe. WR\,125 has a much longer period and lower eccentricity, thus begging the question of its evolutionary status. If the stellar masses were similar to those of WR\,140, then even with the larger distance to WR\,125, the system would be resolvable with the CHARA Array, as \citet{2021ApJ...908L...3R} resolved the WN+O binary WR\,133 with an apparent semi-major axis of less than one milliarcsecond. With a new higher-sensitivity beam combiner coming online \citep{2022SPIE12183E..0NL}, the visual orbit of the fainter WR\,125 may be possible to track in the coming decade.

Another context in which WR\,125 resembled WR\,140 was its non-thermal radio emission pointing to the presence of shock-accelerated electrons in its wind \citep{1986ApJ...303..239A}. The fading of this \citep{1992MNRAS.258..461W}, like that of WR\,140 suggesting the burial of the non-thermal source in the wind \citep{1990MNRAS.243..662W} intensified the infrared monitoring of WR\,125 leading to the discovery of the first dust-formation episode.
While \citet{2019MNRAS.484.2229M} presented a very limited X-ray data set, the similarities of WR\,125 to WR\,140 demands further analysis with the addition of more data. In particular, the X-ray light curve of WR\,140 is explained through the radiative transfer based on hydrodynamical simulations based on the orbital geometry \citep{2021ApJ...923..191P}. With additional data from X-ray satellites and the orbit we presented in Section \ref{sec:orbit}, similar computations may allow for a complete understanding of the gas and its cooling and dust formation for WR\,125. Every astrophysical laboratory in the population of dust-forming WC binaries allows us additional information into the physics of dust formation in the extreme environments surrounding the WCd binaries.

The infrared light curve shows a remarkable repeatability in its shape between the two observed outbursts, as seen in the phased light curves in Fig.~\ref{fig:Phase}. Our observations with SOFIA show a cooler dust cloud than seen earlier in the dust outburst with Subaru/COMICS \citep{2022ApJ...930..116E}. As our observations were taken after the near-infrared peak in the light curve and the observations of \citet{2022ApJ...930..116E} were taken prior to the peak, this is not surprising and supports a dust cloud that forms and then expands and cools as it moves into the interstellar medium. 

The contrast of the dusty spectral features such as the 8 $\upmu$m and 6.2 $\upmu$m features compared to the continuum and even in comparison to the WR emission lines appears to be less than the earlier observations taken in 2019. This shows that the continuum emission processes are dominant by the time the dust begins to cool in these systems. Likely, the changes in the 6.2 $\upmu$m emission line seen in the early phases of the dust outburst of WR\,137 \citep{2023arXiv230811798P} when combined with the low contrast seen in these later phases of dust production for WR\,125 give us a means to fully understand the chemistry of dust production in these binaries. Namely, we must be able to observe the spectral features at high signal-to-noise at several epochs early in the dust formation episodes in order to model the dust production and understand the chemistry of these systems. Given the potential importance of WC dust in our understanding of the modern-day and early-Universe dust budget \citep[e.g.][]{2017MNRAS.468.2416M,2020ApJ...898...74L}, such observations and the associated laboratory measurements are crucial to understand the relative importance of WC dust.

Lastly, we can use the SOFIA observations and contemporaneous infrared photometry to estimate the mass of the dust produced in this binary system. From the G21 analysis shown in Fig.~\ref{fig:blackbody}, we first note a few parts of this analysis. The fluxes  at the lowest and highest wavelengths are seen to lie above the blackbody fit to the spectrum shown that has a temperature of $\sim$400 K.
We divided the residual spectra shown in Fig.~6 by the emissivity for amorphous carbon grains \citep[][ACAR sample]{1996MNRAS.282.1321Z}. Using all the IR data and the dust emissivity, we can best fit the distribution with a temperature of 500 K, which is consistent with the temperature fitted to the multi-wavelength IR photometry of \citet{1994MNRAS.266..247W} in 1993, at a similar orbital phase to the SOFIA observations. The dust temperature is a reasonable fit over 2--20 $\upmu$m, but deviates towards the short and long wavelength regions of the range, indicating that the data are not well represented with a single dust temperature. As the dust has been moving away from the star for roughly three years since the beginning of the present dust-formation episode at this time and the stellar radiation heating it diluted, there will be a range of dust temperatures, with the oldest dust being coolest. Fitting multiple components to the flux distribution at this time is beyond the scope of this paper.

If we adopt an isothermal model for the dust with T = 500 K, a distance to the system of 5.88 kpc \citep{2021AJ....161..147B}, we get a dust mass of $1.5 \times 10^{-6} M_\odot$ created, a bit higher than that inferred from the data in 1993 reported by \citet{1994MNRAS.266..247W} but with a similar temperature. {If the temperature is increased from 500 K to 550 K, then the long wavelength fluxes no longer fit the blackbody distribution, whereas if we decrease the temperature from 500 K to 450 K, the short wavelength fluxes are no longer fit, so we estimate an error of the temperature for an {\it isothermal} model to be $\sim25$ K.} With the $\sim$3-year dust formation prior to our observation, we can then estimate the amount of carbon available for dust formation.  For this, we can consider the mass-loss parameters of  \citet{2012A&A...540A.144S}, where we find that $\dot{M} = 3.05 \times 10^{-5} M_\odot {\rm yr}^{-1}$ and a carbon fraction of 0.4. We estimate that $\sim$10\% of the WC wind enters the shock region based on an opening angle of 35--40$^\circ$ dependent on
the wind parameters of the two stars. From these considerations, we would see that the available carbon that had gone into the shock by the time
of our SOFIA observation is about $3.5 \times 10^{-6} M_\odot$, with about a quarter of this being condensed into dust. As a comparison, if the dust temperature was 400 K, the dust mass becomes $1.1 \times 10^{-5} M_\odot$, which is higher than the amount of available carbon, further pointing to a need for multiple temperatures in the dust formation region years after periastron. While this seems high at first glance as WR\,140 produces about $2-6 \times 10^{-8} M_\odot$ each periastron \citep{2009MNRAS.395.1749W}, WR\,140's dust production happens over a very small time window in comparison to that of WR\,125. Furthermore, we note that planned observations in the coming years will solve this much better with
high spatial resolution than with our spatially unresolved spectra presented here.

\section{Conclusions} 

Our analysis of WR\,125 has provided several key insights to the system. These include:
\begin{itemize}
    \item The infrared light curve has been fully consistent across two separate outbursts near the periastron passage of the system. The repeatability is reminiscent of the prototype of the dusty WC binaries: WR\,140. A phase-dispersion minimization routine allowed us to measure the period of the binary to be 28.12$^{+0.10}_{-0.05}$ years. 
    \item From archival spectroscopy spanning $\sim$20 years, we have measured a first orbit of WR\,125. Only the WR star had measurable motion with our data. The eccentricity of the binary is 0.29$\pm$0.12 based on an analysis with the periastron passage held constant between the rise of the NEOWISE-R mid-infrared flux and the peak of the $K-$band near-infrared flux. The O star kinematics was not able to be measured from our spectra, which did not cover the range of the higher Balmer absorption lines.
    \item The infrared spectrum from SOFIA shows a weak signature of a 6.2 $\upmu$m UIR band, similar to that of WR\,137. Unlike the recent analysis of WR\,137, and similar to the results of \citet{1994MNRAS.266..247W}, no Wolf-Rayet emission lines are seen in the mid-infrared when the dust emission is near maximum for this system. Our analysis of the dust emission shows that it cooled from 800 K in 2019 to 580 K in 2021, likely consistent with the dust cloud expanding and cooling as it is radiatively driven away from the binary. Future studies detailing the expansion of the dust will allow this to be fully tested.
    \item Our data support the idea of a non-isothermal dust around WR\,125, although the best fit temperature seems to be near 580 K, resulting in a dust mass of $9.5 \times 10^{-7} M_\odot$ from the recent periastron. 
\end{itemize}

The current outburst of WR\,125 highlights the importance of long-term spectroscopic monitoring of such systems. The next few years will be especially crucial for WR\,125, including the radial velocity maximum for the WR star. Likely, the orbit will be crucial to modeling efforts for any dust images that are made with infrared imaging with either ground-based telescopes or JWST. Long-term monitoring of WCd stars with ground-based photometry is instrumental in determining orbital periods in studies like these, and with a large sample of such systems, we could create an accurate assessment of the amount of dust created in such systems, which could be of cosmological importance. 

\begin{acknowledgments}

Based in part on observations obtained at the international Gemini Observatory with programs GN-2019B-Q-410 and GN-2020A-Q-309 and then processed using the Gemini IRAF package, a program of NSF’s NOIRLab, which is managed by the Association of Universities for Research in Astronomy (AURA) under a cooperative agreement with the National Science Foundation on behalf of the Gemini Observatory partnership: the National Science Foundation (United States), National Research Council (Canada), Agencia Nacional de Investigaci\'{o}n y Desarrollo (Chile), Ministerio de Ciencia, Tecnolog\'{i}a e Innovaci\'{o}n (Argentina), Minist\'{e}rio da Ci\^{e}ncia, Tecnologia, Inova\c{c}\~{o}es e Comunica\c{c}\~{o}es (Brazil), and Korea Astronomy and Space Science Institute (Republic of Korea). This work was enabled by observations made from the Gemini North telescope, located within the Maunakea Science Reserve and adjacent to the summit of Maunakea. We are grateful for the privilege of observing the Universe from a place that is unique in both its astronomical quality and its cultural significance.

Some of the data presented herein were obtained at the W. M. Keck Observatory, which is operated as a scientific partnership among the California Institute of Technology, the University of California and the National Aeronautics and Space Administration. The Observatory was made possible by the generous financial support of the W. M. Keck Foundation.
The authors wish to recognize and acknowledge the very significant cultural role and reverence that the summit of Maunakea has always had within the indigenous Hawaiian community.  We are most fortunate to have the opportunity to conduct observations from this mountain.

Based in part on observations made with the NASA/DLR Stratospheric Observatory for Infrared Astronomy (SOFIA). SOFIA was jointly operated by the Universities Space Research Association, Inc. (USRA), under NASA contract NNA17BF53C, and the Deutsches SOFIA Institut (DSI) under DLR contract 50 OK 2002 to the University of Stuttgart. 

Financial support for this work was provided in part by NASA through award \#08-0220 issued by USRA. ARD is grateful to Embry-Riddle Aeronautical University's Undergraduate Research Institute for financial support through their IGNITE program and the NASA Space Grant program. NDR is grateful for support from the Cottrell Scholar Award \#CS-CSA-2023-143 sponsored by the Research Corporation for Science Advancement as well as support from NASA through award \#09-0163 issued by USRA. 

\end{acknowledgments}

%





\bibliography{sample631}
\bibliographystyle{aasjournal}



\end{document}